\documentclass{PoS}

\PoS{PoS(LAT2005)089}

\usepackage{graphicx}

\usepackage{epsfig}
\usepackage{amsfonts}

\def\half{{\scriptstyle \raise.2ex\hbox{${1\over2}$}}}
\def\fourth{{\scriptstyle \raise.2ex\hbox{${1\over4}$}}}
\def\simle{
    \mathrel{\rlap{\raise 0.511ex 
        \hbox{$<$}}{\lower 0.511ex \hbox{$\sim$}}}}

\newcommand*{\chpt}{\raise0.4ex\hbox{$\chi$}PT}
\newcommand*{\schpt}{S\raise0.4ex\hbox{$\chi$}PT}

\newcommand*{\Tr}{\textrm{Tr}}

\newcommand{\timestwo}{\!\times\!}

\newcommand{\cL}{\mathcal{L}}
\newcommand{\cM}{\mathcal{M}}

\newcommand{\cV}{\mathcal{V}}

\newcommand{\gtwo}{\ensuremath{g\!-\!2}}

\def\eq#1{Eq.~(\ref{eq:#1})}
\def\eqs#1#2{Eqs.~(\ref{eq:#1}) and (\ref{eq:#2})}

\title{Lowest Order Hadronic Contribution to the Muon \gtwo }
\ShortTitle{Lowest Order Hadronic Contribution to the Muon \gtwo }

\author{\speaker{Christopher Aubin}\\
        Physics Department,
        Columbia University, 
        New York, NY 10027, USA\\
        E-mail: \email{caubin@phys.columbia.edu}}
\author{Tom Blum\\
        Department of Physics, 
		University of Connecticut, 
		Storrs, CT 06269, USA\\
        E-mail: \email{tblum@phys.uconn.edu}}

\abstract{
We present the most recent lattice results for the lowest-order hadronic contribution to the muon anomalous magnetic moment using
$2\!+\!1$ flavor improved staggered fermions. A precise fit to the low-$q^2$ region of the vacuum polarization is necessary to accurately extract the muon \gtwo. To obtain this fit, we use staggered chiral perturbation theory with the inclusion of the vector particles as resonances, to evaluate the vacuum polarization. We discuss the preliminary fit results and attendant systematic uncertainties, paying particular attention to the relative contributions of the pions and vector mesons.
}

\FullConference{XXIIIrd 
International Symposium on Lattice Field Theory\\
25-30 July 2005\\
Trinity College, Dublin, Ireland}

\begin{document}

\bibliographystyle{JHEP}

Currently, experimental measurements of the muon \gtwo\ have reached a level of high precision \cite{Bennett:2004pv}. As a check on the current standard model of particle physics, a competitively precise theoretical estimate for this quantity is needed. The current theoretical calculations come from dispersion relations relating the photon vacuum polarization to the cross-section for either $e^+ e^-\to$hadrons or, by using isospin symmetry, $\tau$ decay to hadrons. The agreement between the theoretical calculations coming from $e^+ e^-$ data to experiment is within 2.7 standard deviations, and this is lowered to 1.4 standard deviations if one includes the results incorporating $\tau$ decay \cite{Davier:2003pw, Ghozzi:2003yn}. The need for a determination completely from first principles ({\it i.e.}, using the lattice) is desirable to check the current discrepancy between theory and experiment.

We can extract the muon anomalous magnetic moment by calculating the full muon-photon vertex, where the correction to the tree-level result is given by $a_\mu \equiv (g-2)/2 = F_2(q^2=0)$, where $F_2(q^2)$ is the form factor proportional to $[\gamma^\mu, 
\gamma^\nu]$ in the full vertex, and $q$ is the momentum transferred to the photon. This can be calculated in perturbation theory in the electroweak theory, with the first contribution coming in at $O(\alpha)$, where $\alpha=e^2/4\pi$ is the fine-structure constant. 

\begin{figure}[tp]
\begin{center}
\includegraphics{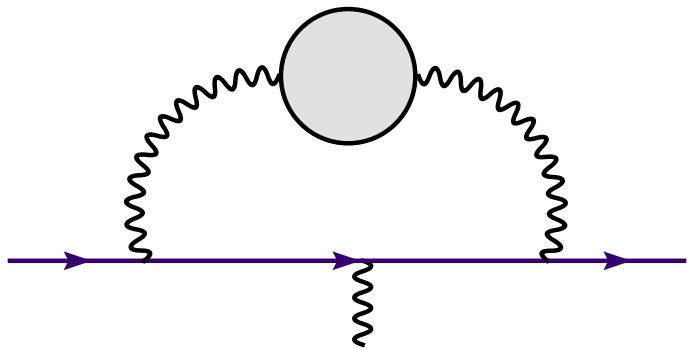}
\caption{The lowest order hadronic contribution to the muon \gtwo. The blob represents all QCD diagrams.}
\label{fig:hadcont}
\end{center}
\end{figure}

The first effects from QCD come in at $O(\alpha^2)$ and are shown in Fig.~\ref{fig:hadcont}. The hadronic contributions are $7\timestwo 10^{-5}$ smaller than the leading corrections and account for most of the discrepancy with experiment. Due to the strength of the strong coupling constant, these terms cannot be reliably calculated in perturbation theory at the relevant energies, so other methods must be employed to calculate them, and this is precisely where the lattice can be useful. 
Currently, the next hadronic contribution, coming from light-by-light scattering at $O(\alpha^3)$, is also being studied using lattice techniques \cite{Hayakawa}.

Lattice QCD provides a first principles method to calculate the hadronic contribution to the muon \gtwo. We can calculate the QCD contribution to the muon \gtwo\ coming from the vacuum polarization of the photon using Refs.~\cite{g-2_1,g-2_2,g-2_3},
\begin{equation}
	a_\mu^{(2){\rm had}} = 
	\left(\frac{\alpha}{\pi}\right)^2
	\int_0^\infty dK^2 f(K^2)\Pi(K^2)\ ,
\end{equation}
where $\Pi(K^2)$ is the vacuum polarization of the photon, which we calculate on the lattice, and $f(K^2)$ is a kernel given in Ref.~\cite{g-2_1}. Since $f(K^2)$ diverges as $K^2\to 0$, we must be able to calculate the low momentum region of the vacuum polarization very precisely, and thus large lattices will be needed here. This is also why perturbation theory is not reliable, since the largest contribution to $a_\mu^{(2){\rm had}}$ comes from the low-energy regime.

A detailed discussion of the lattice calculation can be found in Refs.~\cite{g-2_1,g-2_2}. We used lattices generated by the MILC collaboration using $2\!+\!1$ flavors of light dynamical improved staggered quarks. The results shown here are those coming from three of the ``fine'' lattices ($a\approx 0.086$ fm) with a lattice strange sea quark mass of $0.031$. The light sea quark mass in lattice units is $0.0062$ on the lattice volume of $28^3\timestwo 96$ and $0.0031$ on the $40^3\timestwo 96$ volume. 

In order to fit to the low-momentum region, we use Staggered Chiral Perturbation Theory (\schpt) \cite{SCHPT} coupled to photons. The generalization of the standard staggered chiral Lagrangian to include external photons amounts to changing the derivative operator to a covariant derivative: $\partial_\mu\Sigma \to D_\mu\Sigma = \partial_\mu\Sigma + i e 
A_\mu [Q,\Sigma]$, where $\Sigma$ is the $12\timestwo 12$ matrix 
of pseudo-Goldstone bosons, $Q$ is the light quark charge matrix in units of $e$ and $A_\mu$ is the photon field. Recall the chiral symmetry of \schpt\ is an $SU(12)_L\timestwo SU(12)_R$ symmetry for 3 flavors of staggered quarks, and under this symmetry $\Sigma$ transforms as $\Sigma\to L\Sigma R^\dagger$ with $L\in SU(12)_L$ and $R\in SU(12)_R$.
Our (Euclidean) Lagrangian in this case is given by
\begin{equation}
  	\cL  = 
	\frac{f^2}{8}\Tr[D_\mu\Sigma D_\mu\Sigma^\dagger]
  	- \frac{\mu f^2}{4}\Tr[\cM\Sigma + 	
	 	\Sigma^\dagger\cM] + a^2\cV
\end{equation}
where $\cV$ is the taste-symmetry breaking potential for multiple flavors of staggered quarks \cite{SCHPT} and $\cM$ is the light quark mass matrix. No additional terms arise at this order in the Lagrangian which violate the taste symmetry while being coupled to the external photons. Additionally, we drop the additional singlet mass term from Ref.~\cite{SCHPT} as it will not be relevant for the current calculation.

\begin{figure}[tp]
\begin{center}
\includegraphics{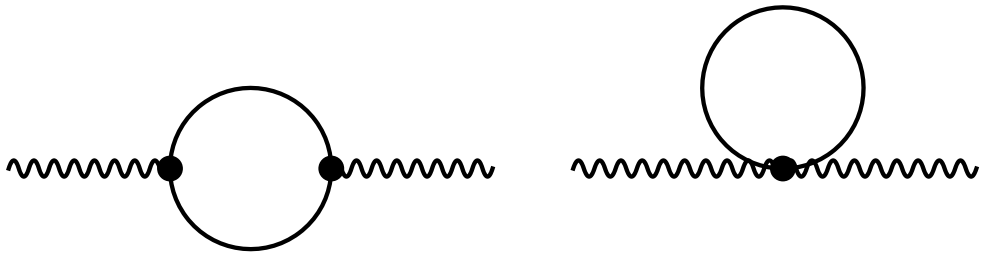}
\caption{The two diagrams that contribute to the photon
vacuum polarization at one loop in \schpt.}
\label{fig:onelooppion}
\end{center}
\end{figure}

Using this, we can calculate the one-loop contribution 
(Fig.~\ref{fig:onelooppion}), which is given by
\begin{equation}\label{eq:onelooppion}
	\Pi(p^2)  =  \frac{\alpha}{4\pi}
      \frac{1}{16}\sum_{i, t} \Biggl\{
   	   \frac{1}{3}\left(1+ 
   	      x_{i_t}\right)^{3/2}
   	    \ln\left( 	  
	    \frac{\sqrt{1+x_{i_t}}+1}{\sqrt{1+x_{i_t}}-1}
       \right)-\frac{2x_{i_t}}{3} 
       - \frac{8}{9} + 
     \frac{1}{3}\ln\left(\frac{m_{i_t}^2}{\Lambda^2}\right)
      \Biggr\} + A
\end{equation}
where $x = 4m^2/p^2$, $i\in\{\pi^+,K^+\}$, and the sum over $t$ is over the 16 tastes of mesons. The masses for these mesons in terms of the Lagrangian parameters are given in Ref.~\cite{SCHPT} and their measured values on the lattice can be found in Ref.~\cite{MILC_SPECTRUM}. This is the only place where the effects of taste violations are seen. There are no neutral mesons in the loops, so we do not get effects from the two-point hairpin insertions \cite{SCHPT}. Additionally, aside from an overall constant, $A$, there are no free parameters in this expression, as all the meson masses have been determined in lattice simulations \cite{MILC_SPECTRUM}. 

The problem, however, is that this is two orders of magnitude too small compared with the lattice data. This indicates that there must be other effects which dominate the low energy physics of the hadronic contribution to the vacuum polarization. We take a lesson from quenched simulations \cite{g-2_3,QCDSF}, where the vacuum polarization can be completely described by the $\rho$. In that case, there is a term coming from the vector bound state and an additional term coming from the two-particle continuum. This implies that perhaps in the dynamical case, there may be large effects coming from vector mesons, despite their large masses, which are not included in standard chiral perturbation theory.

The model we use here is that of the resonance formalism of Ref.~\cite{Ecker:1988te}, where the relevant interaction Lagrangian is
\begin{equation}
	\cL_{\rm vec} = 
	\frac{f_V}{2\sqrt{2}}\Tr\left[V_{\mu\nu}
	(\sigma F_{\mu\nu}\sigma^\dagger
		+ \sigma^\dagger F_{\mu\nu}\sigma)\right]\ .
\end{equation}
$f_V$ is the tree-level vector decay constant, $V_{\mu\nu}$ is the $12\timestwo 12$ matrix of the 9 vector mesons, $\sigma^2=\Sigma$, and $F_{\mu\nu} = eQ(\partial_\mu A_\nu - \partial_\nu A_\mu)$. Although we denote the vector fields using an antisymmetric tensor field, it can be shown that there are still the correct number of degrees of freedom for each particle \cite{Ecker:1988te}. 

Under the chiral $SU(12)_L\timestwo SU(12)_R$ symmetry, we have
\begin{equation}
	V_{\mu\nu} \to U V_{\mu\nu}U^\dagger \ \ ,\ 
	\sigma \to  L\sigma U^\dagger = U\sigma R^\dagger\label{eq:Udef}
\end{equation}
where $U$ is a spacetime-dependent $SU(12)$ matrix defined by this equation.

\begin{figure}[tp]
\begin{center}
\includegraphics[width=\textwidth]{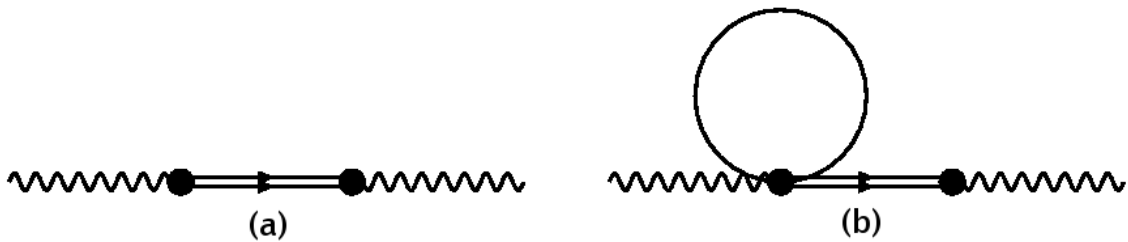}
\caption{The diagrams that contribute to the photon
vacuum polarization with the $\rho$ in \schpt\ at (a) tree level
and (b) one loop. The double line is the vector mesons and the single line is a pion or kaon as before.}
\label{fig:rho}
\end{center}
\end{figure}

The leading order contribution here is the tree-level diagram shown in Fig.~\ref{fig:rho}(a), and is given by
\begin{equation}\label{eq:rho_tree_level}
	\Pi_V(p^2) =
	-\frac{\alpha}{4\pi}\frac{(4\pi f_V)^2}{3}\left[
	\frac{3}{p^2+m^2_{\rho^0}} + \frac{1}{p^2+m^2_{\omega}}\right]\ .
\end{equation}
We have not included the taste index on the vectors, since empirically the taste-breaking for the vectors is negligible \cite{MILC_SPECTRUM}. 
Again, we see that if we measured $f_V$ and the vector masses on the lattice, there are no free parameters in this expression. Also, although the masses of the vectors are much heavier than the pion, there is the enhancement factor of $(4\pi f_V)^2$ in the numerator here, which we can see makes this the dominating factor in the overall result. Ref.~\cite{Becirevic:2003pn} estimates $f_V$ to be about 200 MeV, so this term is roughly $O(1)$ in the chiral power counting scheme, and thus will dominate over the one-loop pion term.

Since we worked to one-loop order in the pion sector, we should do so in the $\rho$ sector, as these should be roughly the same order. The only contributions are tadpole corrections to the $\rho-\gamma$ vertex, Fig.~\ref{fig:rho}(b). We get
\begin{equation}
\Pi_{V}^{\rm 1-loop}(p^2)
	=\frac{\alpha}{4\pi}\left(\frac{4 f^2_V}{ f^2} \right)
	\frac{1}{16}\sum_t \Biggl[
	2\frac{m_{\pi_t}^2 \ln m_{\pi_t}^2}{p^2+m_\rho^2} +
	\frac{m_{K_t}^2 \ln m_{K_t}^2}{p^2+m_\rho^2} +
	\frac{m_{K_t}^2 \ln m_{K_t}^2}{p^2+m_\omega^2}	\Biggr]
\end{equation} 
We can see again that the taste violations here enter quite trivially as they did in the pure pion sector.

\begin{figure}[tp]
\begin{center}
\includegraphics[width=.9\textwidth]{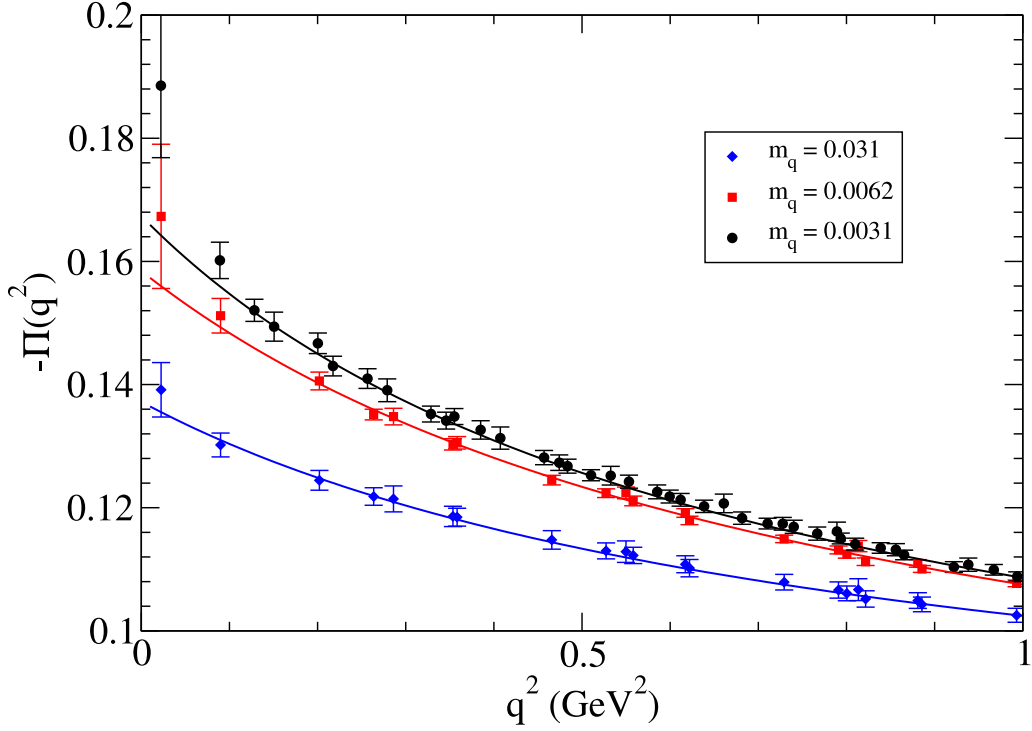}
\caption{Fits to the lattice data using the 
tree-level vector formula for the three different quark masses for the low-$q^2$ region.}
\label{fig:fits}
\end{center}
\end{figure}

Figure \ref{fig:fits} shows the data for the three different masses used with the results from fitting to \eq{rho_tree_level}. Although these fits only include the tree-level $\rho$ results for now, the inclusion of the one-loop formulae changes the fit only slightly. We can see that \eq{rho_tree_level} fits quite well, except for lighter quark masses and the smallest value of $q^2$, where the fit function undershoots the data. There are several reasons for why this may occur. We do not include the Naik term when calculating the quark propagators, though it is hard to see how this would affect the low $q^2$, long-distance, behavior. It is possible the effect is due to finite volume, though in the free theory the vacuum polarization turns over the other way from small $q^2$.
Finally, it may simply be due to poor statistics.


Using the results of these fits, we find when extrapolating to the physical values of the quark masses that the hadronic contribution to the muon \gtwo\ (from the $q^2<1$ GeV regime) is $a_\mu^{\rm had, VP} = 657(20)\timestwo 10^{-10}$. The perturbative contribution coming from the $q^2>1$ GeV regime is $a_\mu^{\rm had, VP, pert} \simle 10\timestwo 10^{-10}$, and the final result is the sum of these two terms. The error quoted here is statistical only. This is somewhat lower than the results from dispersion relations, which is $a_\mu^{\rm had, disp} =  693.4(5.3)(3.5)\timestwo 10^{-10}$. The difference with the previous lattice result of $a_\mu^{\rm had, VP} = 545(65)\timestwo 10^{-10}$ \cite{g-2_3} comes mainly from the inclusion of the $0.0031$ light quark mass result which allows an extrapolation, using \eqs{onelooppion}{rho_tree_level}, to the physical quark mass. This accentuates the fact that a theoretically motivated fitting function is necessary to obtain an accurate determination of the low-$q^2$ region. We also note that the final value of the lowest order hadronic correction may increase further if
the observed overshoot of the fit function by the lattice data turns out to be physical. Finally, we caution that the disconnected diagrams (connected by gluons) that contribute to the lowest order vacuum polarization have not been included yet. These are Zweig suppressed in general and vanish exactly in the $SU(3)$ flavor limit, but may contribute on the one-to-few percent level.

This work was supported by the U.S.\ DOE. We thank NERSC for the computational resources expended on this project. TB thanks Michael Ramsey Musolf for helpful discussions.


\begin{thebibliography}{10}

\bibitem{Bennett:2004pv}
{\bf Muon g-2} Collaboration, G.~W. Bennett {\em et~al.}, {\it Measurement of
  the negative muon anomalous magnetic moment to 0.7-ppm},  {\em Phys. Rev.
  Lett.} {\bf 92} (2004) 161802,
  [\href{http://xxx.lanl.gov/abs/hep-ex/0401008}{{\tt hep-ex/0401008}}].

\bibitem{Davier:2003pw}
M.~Davier, S.~Eidelman, A.~Hocker, and Z.~Zhang, {\it Updated estimate of the
  muon magnetic moment using revised results from e+ e- annihilation},  {\em
  Eur. Phys. J.} {\bf C31} (2003) 503--510,
  [\href{http://xxx.lanl.gov/abs/hep-ph/0308213}{{\tt hep-ph/0308213}}].

\bibitem{Ghozzi:2003yn}
S.~Ghozzi and F.~Jegerlehner, {\it Isospin violating effects in e+ e- vs. tau
  measurements of the pion form factor |f(pi)|**2(s)},  {\em Phys. Lett.} {\bf
  B583} (2004) 222--230, [\href{http://xxx.lanl.gov/abs/hep-ph/0310181}{{\tt
  hep-ph/0310181}}].

\bibitem{Hayakawa}
M.~Hayakawa {\em et~al.}, {\it Hadronic light-by-light scattering contribution to
  the muon g-2 from lattice QCD: Methodology}, 
\newblock these proceedings.

\bibitem{g-2_1}
T.~Blum, {\it Lattice calculation of the lowest order hadronic contribution to
  the muon anomalous magnetic moment},  {\em Phys. Rev. Lett.} {\bf 91} (2003)
  052001, [\href{http://xxx.lanl.gov/abs/hep-lat/0212018}{{\tt
  hep-lat/0212018}}].

\bibitem{g-2_2}
T.~Blum, {\it Lattice calculation of the lowest order hadronic contribution to
  the muon anomalous magnetic moment: An update with Kogut-Susskind fermions},
  {\em Nucl. Phys. Proc. Suppl.} {\bf 129} (2004) 904--906,
  [\href{http://xxx.lanl.gov/abs/hep-lat/0310064}{{\tt hep-lat/0310064}}].

\bibitem{g-2_3}
T.~Blum, {\it Lattice calculation of the lowest-order hadronic contribution to
  the muon anomalous magnetic moment},  {\em Nucl. Phys. Proc. Suppl.} {\bf
  140} (2005) 311--313, [\href{http://xxx.lanl.gov/abs/hep-lat/0411002}{{\tt
  hep-lat/0411002}}].

\bibitem{SCHPT}
C.~Aubin and C.~Bernard, {\it Pion and kaon masses in staggered chiral
  perturbation theory},  {\em Phys. Rev.} {\bf D68} (2003) 034014,
  [\href{http://xxx.lanl.gov/abs/hep-lat/0304014}{{\tt hep-lat/0304014}}].

\bibitem{MILC_SPECTRUM}
C.~W. Bernard {\em et~al.}, {\it The QCD spectrum with three quark flavors},
  {\em Phys. Rev.} {\bf D64} (2001) 054506,
  [\href{http://xxx.lanl.gov/abs/hep-lat/0104002}{{\tt hep-lat/0104002}}].

\bibitem{QCDSF}
{\bf QCDSF} Collaboration, M.~Gockeler {\em et~al.}, {\it Vacuum polarisation
  and hadronic contribution to muon g-2 from lattice QCD},  {\em Nucl. Phys.}
  {\bf B688} (2004) 135--164,
  [\href{http://xxx.lanl.gov/abs/hep-lat/0312032}{{\tt hep-lat/0312032}}].

\bibitem{Ecker:1988te}
G.~Ecker, J.~Gasser, A.~Pich, and E.~de~Rafael, {\it The role of resonances in
  chiral perturbation theory},  {\em Nucl. Phys.} {\bf B321} (1989) 311.

\bibitem{Becirevic:2003pn}
D.~Becirevic, V.~Lubicz, F.~Mescia, and C.~Tarantino, {\it Coupling of the
  light vector meson to the vector and to the tensor current},  {\em JHEP} {\bf
  05} (2003) 007, [\href{http://xxx.lanl.gov/abs/hep-lat/0301020}{{\tt
  hep-lat/0301020}}].

\end{thebibliography}

\providecommand{\href}[2]{#2}\begingroup\raggedright\endgroup

\end{document}